\documentclass{new_tlp}
\usepackage{url} 
\usepackage{mathptmx}
\usepackage{fontawesome} 
\usepackage[utf8]{inputenc}   
\usepackage{mathtools}  
\usepackage{graphicx} 
\usepackage{changepage} 
\usepackage{latexsym} 
\usepackage{wrapfig} 
\usepackage{mathtools} 
\usepackage{xcolor}
\usepackage{soul}
\usepackage[utf8]{inputenc}
\usepackage[small]{caption}
\usepackage{xcolor} 
\usepackage{amsmath}       
\usepackage{amssymb}            
\usepackage{tikz}  
\usetikzlibrary{arrows.meta}  
\usepackage{tcolorbox} 
\tcbuselibrary{raster,skins}
\usepackage{rotating}    
    
\usepackage{tikz-cd}  
\newcommand{\hide}[1]{}

\newcommand{\ADFg}{\textsf{adfg}}

\newcommand{\FFMMA}{\ensuremath{F^{\MMA}}}
\newcommand{\FFADF}{\ensuremath{F^{\ADF}}}
\newcommand{\FMMA}{\ensuremath{\mathcal{F}^{\MMA}}}
\newcommand{\FADF}{\ensuremath{\mathcal{F}^{\ADF}}}

\newcommand{\exact}{\textsf{exact}} 
\newcommand{\maxi}{\textsf{twoVal}}

\newcommand{\ADF}{\textsf{ADF}}  
\newcommand{\MMA}{\textsf{MMA}}

\newcommand{\Label}{\textsf{L}}

\newcommand{\inL}{\textsf{in}}
\newcommand{\outL}{\textsf{out}}
\newcommand{\forinL}{\textsf{out}}
\newcommand{\foroutL}{\textsf{in}}
\newcommand{\undecL}{\textsf{undec}}

\newcommand{\pre}{\ensuremath{\textsf{pred}}}

\newcommand{\andC}{\textsf{and}}

  \title[{\small Abstract Interpretation in Formal Argumentation}]
        {Abstract Interpretation in Formal Argumentation:  with 
	a Galois Connection for Abstract Dialectical Frameworks 
  and May-Must Argumentation \\ (First Report)}

  \author[R. Arisaka and T. Ito]
         {Ryuta Arisaka and Takayuki Ito\\
         Nagoya Institute of Technology, Nagoya, Japan\\
         \email{ryutaarisaka@gmail.com, ito.takayuki@nitech.ac.jp}}
\jdate{March 2003}
\pubyear{2003}
\pagerange{\pageref{firstpage}--\pageref{lastpage}}
\doi{S1471068401001193}

\newtheorem{definition}{Definition}[section] 
\newtheorem{example}{Example}[section]  
\newtheorem{lemma}{Lemma}[section] 
\newtheorem{proposition}{Proposition}[section]
 
\newtheorem{theorem}{Theorem}[section]   
\newtheorem{corollary}{Corollary}[section]    

\begin{document}
\nocite{*}

\label{firstpage}

\maketitle

  \begin{abstract}    
	  Labelling-based formal argumentation 
	  relies on labelling functions that typically assign 
	  one of 3 labels to indicate either acceptance, rejection, 
	  or else undecided-to-be-either, to each argument. While 
	  a classical labelling-based approach applies 
	  globally uniform conditions as to how an argument is to be 
	  labelled,  they can be determined more locally 
	  per argument. 
	  Abstract dialectical frameworks (\ADF) is a well-known 
	  argumentation formalism that belongs to this category, 
	  offering a greater labelling flexibility. 
	  As the size of an argumentation 
	  increases in the numbers of arguments and argument-to-argument
	  relations, however, it becomes increasingly more costly    
	  to check whether a labelling function satisfies 
	  those local conditions or even whether the conditions 
	  are as per the intention of those who had specified 
	  them. Some compromise is thus required  
	  for reasoning about a larger argumentation. 
	  In this context, there is 
	  a more recently proposed  formalism of may-must argumentation 
	  (\MMA) that enforces still local but more abstract 
	  labelling conditions. We identify how they link to each other in 
	  this work. 
	  We prove that there is a Galois 
	  connection between them, in which 
	  {\ADF} is a concretisation of {\MMA} and 
	  {\MMA} is an abstraction of \ADF. 
	  We explore  
	  the consequence of 
	  {\it abstract interpretation at play} in 
	  formal argumentation, demonstrating a sound reasoning 
	  about the judgement of acceptability/rejectability in  
	  {\ADF} from within 
	   \MMA.   
	  As far as we are aware, there is seldom any work 
	  that incorporates abstract interpretation 
	  into formal argumentation in the literature, and, in the stated context,
	  this work is the first 
	  to demonstrate its use and relevance. 
  \end{abstract}


\section{Introduction}    
Abstract argumentation \cite{Dung95} represents an argumentation 
as a directed graph of: nodes for arguments; and edges for  
attacks from the source arguments to the target arguments, with an intent to 
infer acceptance statuses of arguments. 
Classically \cite{Dung95,Jakobovits99,Caminada06}, conditions 
for acceptance and rejection are defined 
globally uniformly. However, 
it is also possible to localise the conditions 
to a sub-argumentation 
or even to each argument. 
Abstract Dialectical Frameworks (\ADF) \cite{Brewka13} and  
May-Must Argumentation (\MMA) \cite{arisaka2020a} both 
belong to the last category. 

In \ADF, for each argument, one of 3 acceptance statuses: $\inL$ to mean accepted; 
$\outL$ to mean rejected; 
and $\undecL$ to mean undecided-to-be-either, 
is chosen based on acceptance statuses of 
the arguments attacking. \footnote{{\ADF} actually 
keeps the argument-to-argument relation to be of an unspecified 
nature, allowing in particular the support 
relation to be expressed; however, in the present 
paper, we will abstract any non-attack 
relations. 
Also, it uses 
\textsf{t} for \textsf{in} and \textsf{f} for \textsf{out}; for this, 
however, we use more widely-used notations \cite{Jakobovits99,Caminada06}.} 
In effect, to each argument $a$ attacked by $n$ arguments 
$a_1, \ldots, a_n$, {\ADF} attaches 
 instructions: ``If acceptance 
statuses of $(a_1, \ldots, a_n)$ are $([x_1], \ldots, [x_n])$, then  
choose $[x]$ for $a$'s acceptance status'', 
where every $[\cdot]$ is either of $\inL$, $\outL$, and $\undecL$, 
with $[x_1], \ldots, [x_n]$ covering all combinations of the 3 statuses 
(and 
thus there can be up to $3^n$ instructions).  
As a small example, suppose an argumentation graph 
of: 
\begin{tikzcd}[
  column sep=small, row sep=small,inner sep=0pt] 
     a_p \arrow[r] & a_r \arrow[l,shift left] \arrow[r] & a_q \arrow[l, shift left] 
\end{tikzcd} 
with 3 arguments, where $a_p$ is given:  
``If (acceptance status of) $a_r$ is  $\inL$, 
then choose $[x_1]$ for $a_p$('s acceptance 
status)''; ``If $a_r$ is  
$\outL$, then choose $[x_2]$ for $a_p$''; and ``If $a_r$ is  
$\undecL$, then choose $[x_3]$ for $a_p$''. Similarly 3 cases are considered for $a_q$.  
Meanwhile, instructions for $a_r$ have to cover $9$ cases; e.g. 
``If $(a_p, a_q)$ is: 
\begin{itemize} 
	\item  $(\undecL, \undecL)$ or $(\undecL, \outL)$, then 
		choose $\undecL$ for $a_r$. 
        \item  $(\outL, \undecL)$ or $(\outL, \outL)$, then  
		choose $\inL$ for $a_r$.  
	\item  $(\inL, \undecL)$ or $(\undecL, \inL)$ or $(\inL, \inL)$, 
		then choose $\outL$ for $a_r$. 
	\item   $(\inL, \outL)$ or $(\outL, \inL)$, then choose $\undecL$  for $a_r$.''
\end{itemize}  
\subsection{Research problems} 
With \ADF, a user can specify 
argument's acceptability status independently for each combination. The 
freedom, however, comes with an associated cost.  
Given the complexity results \cite{Brewka13}, 
  increases in the numbers of arguments and 
 argument-to-argument relations can make it 
exponentially more costly to check whether an 
acceptance status of an argument 
satisfies the instructions attached to it, or even whether 
the instructions are as per the intention of those who had specified them. 
For scalability of reasoning about a larger argumentation, techniques 
of search space reduction ought to be 
explored.

\subsection{Contributions} 
We identify in this paper that {\MMA} \cite{arisaka2020a} 
serves to 
abstract 
\ADF's instructions. 
We prove that there is in fact a Galois connection 
for the two formalisms,\footnote{A Galois connection 
for two systems is a pair of mappings betwen them; 
formal detail is in Section 2.} in which, on one hand, {\ADF} is a concretisation 
of {\MMA} and, on the other hand, {\MMA} is an abstraction of \ADF. 
The consequence, as we will show, is abstract interpretation \cite{Cousot77,Cousot79} 
enabling reasonings 
about acceptance statuses of arguments in {\ADF} from within \MMA.

To give contexts to our idea, 
in \MMA, every argument $a$ in an argumentation graph is assigned two 
pairs of natural numbers. One pair, say $(n_1, n_2)$, states 
that at least $n_1$ (resp. $n_2$) arguments attacking $a$ need to be 
rejected for $a$ to be judged possibly accepted (resp. accepted).  
Another pair, say $(m_1, m_2)$, states that at least $m_1$ (resp. $m_2$) 
arguments attacking $a$ need to be accepted for $a$ to be 
judged possibly rejected (resp. rejected). 
As we see, these conditions only specify the cardinality 
of rejected or accepted attacking arguments and not 
specifically which ones. Moreover, 
while $n_1, n_2, ,m_1, m_2$ specify the minimum numbers 
for the respective conditions, any numbers that exceed them 
satisfy the same conditions, that is, these conditions are monotonic. 
Along with any other differences, the level of 
abstractness is thus higher for 
{\MMA} than \ADF. 
Suppose the following 
for the above example 
\begin{tikzcd}[
  column sep=small, row sep=small,inner sep=0pt]  
  \underset{((n^{a_p}_1, n^{a_p}_2), (m^{a_p}_1, m^{a_p}_2))}{a_p}  
	\arrow[r] & 
  \underset{((n^{a_r}_1, n^{a_r}_2), (m^{a_r}_1, m^{a_r}_2))}{a_r}
	\arrow[l,shift left] \arrow[r] & 
	\underset{((n^{a_q}_1, n^{a_q}_2), (m^{a_q}_1, m^{a_q}_2))}{a_q} \arrow[l,shift left] , 
\end{tikzcd}  
with associated pairs, then $((n_1^{a_r}, n_2^{a_r}), (m_1^{a_r}, m_2^{a_r})) = ((1, 2), (1,1))$ 
signifies the following: 
\begin{itemize}
	\item $a_r$ is not judged accepted or rejected in any degree 
		 if neither 
		$a_p$ nor $a_q$ is $\inL$ or $\outL$. 
		According to \cite{arisaka2020a}  
		(Section 2 of this paper for more detail), this case results 
		in the choice of $\undecL$ for $a_r$. 
	\item $a_r$ is judged only possibly accepted, and 
		not judged rejected in any degree if one of $a_p$ and $a_q$ 
		is $\outL$ and the other is $\undecL$. 
                This case results 
		in the choice of either $\inL$ or $\undecL$ for $a_r$ 
		on a non-deterministic basis. 
        \item $a_r$ is judged accepted but not rejected in 
		any degree,  
		  if both $a_p$ and $a_q$ are $\outL$.  
		  This case results 
		in the choice of $\inL$ for $a_r$.   
	\item $a_r$ is not judged accepted in any degree, but judged 
		rejected if both $a_p$ and $a_q$ are $\inL$.  
		This case results 
		in the choice of $\outL$ for $a_r$.  
	\item $a_r$ is judged only possibly accepted and 
		judged rejected if one of $a_p$ and $a_q$ 
		is $\outL$ and the other is $\inL$. 
                This case results 
		in the choice of either $\outL$ or $\undecL$ for $a_r$ 
		on  a non-deterministic basis. 
	\item (The other cases do not happen with the chosen 
		$n_1^{a_r}, n_2^{a_r}, m_1^{a_r}, m_2^{a_r}$.) 
	\end{itemize}
There is the following correspondence between {\ADF} and {\MMA} for 
this example, for any acceptance statuses of $(a_p, a_q)$. 
\begin{itemize} 
	\item When there is only one acceptance status to choose 
		for 
		$a_r$ in this {\MMA} for the acceptance statuses of 
		$(a_p, a_q)$, 
             then the same acceptance status is chosen for $a_r$ in \ADF. 
     \item When there are more than one acceptance status to choose 
	     for $a_r$ in 
		 this {\MMA} for the acceptance statuses of 
		 $(a_p, a_q)$, then one 
		 of them is chosen for $a_r$ in \ADF. 
\end{itemize} 
In other words, this {\MMA} soundly over-approximates 
the {\ADF} instructions. 

This kind of a technique to reason about 
a system from within its abstraction, in a 
manner ensuring that some properties of the abstracted system 
be sound over-approximations of some properties of the concrete system, 
is known as abstract interpretation \cite{Cousot77,Cousot79}, 
which is popular in static program analysis. It is 
almost not studied in formal argumentation, however. Perhaps, 
to the static analysis community, it is a question just 
what of formal argumentation may require abstract interpretation; 
and, for the formal argumentation community, its focus 
having been more on making the prediction of abstract argumentation 
\cite{Dung95} increasingly more precise may explain why 
the concept of {\it abstract interpretation in formal argumentation} 
has been rather elusive. Nonetheless, we contend that 
a stronger move towards reasoning about a larger argumentation 
is bound to gather force, especially 
with an increasing interest in argumentation mining 
technology to automatically extract large-scale argumentations. 
It is a reasonable projection that  
abstract interpretation will play 
just as important a part in formal argumentation as it does in static analysis. 
We take an initiating step towards the development. 

\subsection{Related work} 
As far as we are aware, there is one preprint 
for loop abstraction \cite{ArisakaDauphin18} that 
takes an inspiration from 
abstract interpretation. However, it gives more weights to 
learning about some otherwise unlearnable acceptance statuses in an 
original argumentation 
than to making generally sound reasoning about the properties 
of the original argumentation.  
For its application in the stated 
context,
there is, to the best of our knowledge, none existing in the literature.\footnote{
Google Scholar Search as of 26th May 2020 had produced only 7 results  
out of 4,300+ papers that cited \cite{Dung95} 
and that included ``abstract interpretation'', of which 
only \cite{ArisakaDauphin18} refers to a Galois connection and 
concrete/abstract spaces.} 
We make clear to what extent 
a Galois connection for {\ADF} and {\MMA} permits us 
the above-mentioned sound reasoning 
about {\ADF} within \MMA. \\

In the rest, we will go through technical preliminaries (in Section 2), 
and establish a Galois connection between 
{\MMA} and {\ADF}, showing how it can be utilised 
for soundly reasoning about {\ADF} properties in a corresponding 
abstract space, within {\MMA} (in Section 3).

\section{Technical Preliminaries}    
\subsection{Syntax of abstract argumentation and labelling}  
Let $\mathcal{A}$ denote a class of abstract entities that 
we understand as arguments, and let $\mathcal{R}$ denote 
a class of all binary relations over $\mathcal{A}$. We 
refer to a member of $\mathcal{A}$ (resp. $\mathcal{R}$) 
by $A$ (resp. $R$) with or without a subscript. 
By $\mathcal{R}^A$ for $A \subseteq \mathcal{A}$, we denote 
a subclass of $\mathcal{R}$ 
which contains all and only members $R$ of $\mathcal{R}$  
over $A$, i.e. for every $R \in \mathcal{R}^A$ and every 
$(a_1, a_2) \in R$, it holds that $a_1, a_2 \in A$.   
A (finite) abstract argumentation is then a tuple 
$(A, R)$ with $A \subseteq_{\text{fin}} \mathcal{A}$  
and $R \in \mathcal{R}^A$. 

For characterisation of acceptance statuses, we will 
make use of labellings \cite{Jakobovits99,Caminada06} 
uniformly  
for a compatibility with {\ADF} \cite{Brewka13} and 
{\MMA} \cite{arisaka2020a}, both of which are to be introduced 
below.   
Readers interested in non-labelling-based  
approaches are referred to \cite{Dung95,Baroni07}.

Let $\mathcal{L}$ denote $\{\inL,\outL,\undecL\}$,
and let $\Lambda$ denote the class of all partial 
functions $\mathcal{A} \rightarrow \mathcal{L}$. 
Let $\Lambda^A$ for $A \subseteq 
\mathcal{A}$ denote a subclass of $\Lambda$ 
that includes all (but nothing else) $\lambda \in \Lambda$ 
that are defined for all members (but nothing else) of $A$.   
For the order among members of $\Lambda$, let $\preceq$ be a binary relation over $\Lambda$  
such that $\lambda_1 \preceq \lambda_2$ 
for $\lambda_1, \lambda_2 \in \Lambda$ iff 
all the following conditions hold. (1) There 
is some $A \subseteq_{\text{fin}}
 \mathcal{A}$ such that $\lambda_1, \lambda_2 \in 
\Lambda^A$. (2) For every $a \in A$, $\lambda_1(a) = \inL$ 
(resp. $\lambda_1(a) = \outL$) 
materially implies $\lambda_2(a) = \inL$ (resp. 
$\lambda_2(a) = \outL$). We may write 
$\lambda_1 \prec \lambda_2$ when 
$\lambda_1 \preceq \lambda_2$ but not 
$\lambda_2 \preceq \lambda_1$.  

\subsection{{\MMA} and labelling instructions}    
Let 
$\mathbb{N}$ be the class of natural numbers 
    including 0,  
    and 
for any tuple $T$ of $n$-components, 
let 
$(T)^i$ for $1 \leq i \leq n$ refer to 
$T$'s $i$-th component. 
    A {\it may-must scale} 
    is some 
    $\pmb{X} \in \mathbb{N} \times \mathbb{N}$ 
    with $(\pmb{X})^1 \leq (\pmb{X})^2$. 
    $(\pmb{X})^1$ (resp. $(\pmb{X})^2$) is 
    called {\it may- condition} (resp. {\it must- condition}) 
   of $\pmb{X}$. A {\it nuance tuple} is 
    a pair 
   $(\pmb{X}_1, \pmb{X}_2)$ of may-must scales, 
   the first one $\pmb{X}_1$ for acceptance judgement and the second 
   one $\pmb{X}_2$ for rejection judgement. Cf. Section 1. 
   With $Q \equiv ((n_1, n_2), (m_1, m_2))$, 
   $(Q)^1 = (n_1, n_2)$, $(Q)^2 = (m_1, m_2)$, 
   $((Q)^1)^i = n_i$, and $((Q)^2)^i = m_i$ ($i \in \{1,2\}$). 
   We denote 
   the class of all nuance tuples by $\mathcal{Q}$ and refer 
   to its member by $Q$ with or without a subscript.  
   For any $Q
   \in \mathcal{Q}$, 
   we call $(Q)^1$ its may-must acceptance scale and 
   $(Q)^2$ its may-must rejection scale. 
A {\MMA} is then a tuple $(A, R, f_Q)$ with $A \subseteq_{\text{fin}} \mathcal{A}$;
$R \in \mathcal{R}^A$; and $f_Q: A \rightarrow \mathcal{Q}$.   
We denote the class of all {\MMA} tuples by $\mathcal{F}^{\MMA}$, 
and refer to its member by $F^{\MMA}$ with or without a subscript.

In \MMA, acceptance and rejection 
of an argument 
are independently considered, which are later combined into a final 
decision. 
For the independent judgement, 
for any $\FFMMA \equiv (A, R, f_Q)\ (\in \FMMA)$, 
any $a \in A$ and any $\lambda \in \Lambda$, let 
$\pre^{\FFMMA}(a)$ be the set of all $a_x \in A$ 
with $(a_x, a) \in R$, let  
$\pre^{\FFMMA}_{\lambda, \forinL}(a)$ be the set of 
all $a_x \in \pre^{\FFMMA}(a)$ such that 
$\lambda$ is defined for $a_x$ and that: $\lambda(a_x) = \inL$ 
if $a_x$ attacks , 
and let $\pre^{\FFMMA}_{\lambda, \foroutL}(a)$ be the set of 
all $a_x \in \pre^{\FFMMA}(a)$ such that 
$\lambda$ is defined for $a_x$ and that $\lambda(a_x)= \outL$, then  
 $a$ is said to satisfy: 
\begin{itemize} 
	\item may-a(cceptance condition) (resp. may-r(ejection condition)) 
		under $\lambda$ in $\FFMMA$ iff\linebreak 
		$((f_Q(a))^1)^1 \leq |\pre^{\FFMMA}_{\lambda, \outL}(a)|$ 
		(resp. $((f_Q(a))^2)^1 \leq |\pre^{\FFMMA}_{\lambda, \inL}(a)|$).   
	\item must-a(cceptance condition) (resp. must-r(ejection condition)) 
	     under $\lambda$ in $\FFMMA$ iff \linebreak
		$((f_Q(a))^1)^2 \leq |\pre^{\FFMMA}_{\lambda, \outL}(a)|$ 
		(resp. $((f_Q(a))^2)^2 \leq |\pre^{\FFMMA}_{\lambda, \inL}(a)|$).   
	\item may$_s$-a(cceptance condition) (resp. may$_s$-r(ejection 
		condition)) under $\lambda$ in $\FFMMA$ iff \linebreak 
	      $((f_Q(a))^1)^1 \leq |\pre^{\FFMMA}_{\lambda, \outL}(a)| < 
	((f_Q(a))^1)^2$ 
	(resp. $((f_Q(a))^2)^1 \leq |\pre^{\FFMMA}_{\lambda, \inL}(a)| < 
	((f_Q(a))^2)^2$).   
\item not-a(cceptance condition) (resp. not-r(ejection condition))  
	under $\lambda$ in $\FFMMA$ iff \linebreak 
	      $|\pre^{\FFMMA}_{\lambda, \outL}(a)| < 
	((f_Q(a))^1)^1$ 
	(resp. $|\pre^{\FFMMA}_{\lambda, \inL}(a)| < 
	((f_Q(a))^2)^1$).   
\end{itemize} 
Clearly, the may-must conditions are monotonic  
  over the increase in the number of rejected/accepted 
  attacking arguments. 
  If obvious from the context, 
  ``under $\lambda$ in $\FFMMA$'' may be omitted.    
 
  {\MMA} labelling instructions that we saw in Section 1  
  are technically termed {\it label designations} in 
  \cite{arisaka2020a}. They are derived from 
  combining  
  these independent judgements. 
  Specifically, 
  for any $\FFMMA \equiv 
   (A, R, f_Q)\ (\in \FMMA)$, any $a \in A$, and 
  any $\lambda \in \Lambda$, 
  $\lambda$ is said to designate $l \in \mathcal{L}$ 
  for $a$ 
  iff all the following conditions hold. (Cf. Fig. \ref{fig_2} 
  for which label(s) may be designated for each combination.) \\
  \begin{wrapfigure}[11]{r}{5.3cm}  
\vspace{-\intextsep}
 \begin{tikzcd}[column sep=tiny,row sep=tiny,
	 /tikz/execute at end picture={
		 \draw (-1.35, -1.5) -- (-1.35, 1.2);
		 \draw (-2.7, 0.77) -- (2.6, 0.77);}]  
	 & \text{must}\text{-r} & \text{may}_s\text{-r} & \text{not}\text{-r} \\
	  \text{must}\text{-a} & \undecL & \inL ? & \inL \\ 
	 \text{may}_s\text{-a}	 &  \outL ? & \textsf{any} & \inL ? \\
	 \text{not-a} & \outL & \outL ? & \undecL
\end{tikzcd}   
	\caption{Label designation for each combination 
	  of satisfied may-must conditions under a given $\lambda \in \Lambda$. \textsf{any} is any of 
	$\inL$, $\outL$, $\undecL$. \textsf{in}? is any of 
	$\inL$, $\undecL$. $\outL$? is any of 
	$\outL$, $\undecL$.} 
\label{fig_2} 
\end{wrapfigure} 
  \begin{enumerate}
     \item $\lambda$ is defined for every member of $\pre^{\FFMMA}(a)$.  
     \item \mbox{If $l = \inL$, $a$ satisfies may-a but not must-r 
	     (under $\lambda$)}. 
     \item If $l = \outL$, $a$ satisfies may-r but not must-a. 
     \item If $l = \undecL$, then either of the following holds. 
	     \begin{itemize}
		     \item $a$ satisfies must-a and must-r. 
		     \item $a$ satisfies at least either 
			      may$_s$-a or may$_s$-r. 
		     \item $a$ satisfies not-a and not-r. 
	     \end{itemize}
  \end{enumerate} 
  While we refer a reader to \cite{arisaka2020a} for a slower explanation, these 
  label designations are as the result of \MMA's interpretation 
  of the satisfaction conditions under a possible-world 
  perspective. In the classic (i.e. non-intuitionistic) interpretation of modalities 
(see for example \cite{Garson13}), 
a necessary (resp. possible) proposition is true 
iff it is true 
in every (resp. some) possible world 
accessible from the current world, and  a not possible proposition 
is false in every accessible possible world.   
{\MMA} transposes these to the must- may$_s$- conditions by 
taking acceptance for truth and rejection for falsehood, 
obtainining 
for each $a \in A$ that 
its satisfaction of: must-a (resp. must-r) implies  
acceptance (resp. rejection) of $a$ in every accessible possible 
world; may$_s$-a (resp. may$_s$-r) implies 
acceptance (resp. rejection) of $a$ in some 
accessible possible world; and \mbox{not-a} (resp. not-r)   
implies rejection (resp. acceptance) of $a$ in every accessible 
possible world. (Note the use of ``may$_s$'' instead of ``may'' 
here, once both ``may'' and ``must'' are satisfied, it suffices 
to simply consider ``must''.) Any possible world implying only 
acceptance (resp. rejection) of $a$ is implying 
$\inL$ (resp. $\outL$) for $a$. Any possible world implying both acceptance and rejection 
of $a$ is implying an inconsistent acceptance status, i.e. $\undecL$, for 
$a$ in the 
possible world. In the above definition of label designation, 
$\lambda$ designates any $l \in \mathcal{L}$ for $a \in A$ so long as  
there is some structure of possible worlds one of which 
is the current world and an accessility relation 
such that $l$ is implied for $a$ in at 
least one accessible possible world. \\

\noindent $a$'s label is said to be proper\footnote{In \cite{arisaka2020a}, 
``designated'' instead of ``proper'' is used.} under $\lambda$ iff 
(1) $\lambda$ is defined for $a$, and (2) $\lambda$ designates 
$\lambda(a)$ for $a$. If every $a \in A$'s 
label is proper under $\lambda \in \Lambda^A$, then we call $\lambda$ 
an {\it exact labelling} of $\FFMMA$. 
Suppose 
the following $\FFMMA$ 
\begin{tikzcd}[
  column sep=small, row sep=small,inner sep=0pt]  
  \underset{((2,2), (0,0))}{a_p}  
	\arrow[r] & 
  \underset{((1, 2), (1, 1))}{a_r}
	\arrow[l,shift left] \arrow[r] & 
	\underset{((0,0), (2,2))}{a_q} \arrow[l,shift left]  
\end{tikzcd}  
with associated may-must scales. 
Let $[a_1: l_1, \ldots, a_n:l_n]^{\lambda}$  
for $a_1, \ldots, a_n \in A$ and $l_1, \ldots, l_n \in \mathcal{L}$ 
denote some member of $\Lambda^{\{a_1, \ldots, a_n\}}$ 
with $[a_1:l_1, \ldots,a_n:l_n]^{\lambda}(a_i) = l_i$, then 
both $[a_p:\outL,a_r:\outL,a_q:\inL]^{\lambda}$ and 
\mbox{$[a_p:\outL,a_r:\undecL,a_q:\inL]^{\lambda}$} 
are all exact labellings of $\FFMMA$. ($a_p$ satisfies 
must-r and not-a, and $a_q$ must-a and not-r, irrespective of 
attackers' labels, and therefore $a_r$ satisfies 
must-r and may$_s$-a. There are no other cases.) 

\subsection{{\ADF} and labelling instructions}   
While {\ADF} uses its own set of symbols and terminology different from 
those used in abstract argumentation,
in this paper we keep them consistent with {\MMA} notations.  
A finite {\ADF} is 
a tuple $(A, R, C)$ with: $A \subseteq_{\text{fin}} \mathcal{A}$; 
$R \in \mathcal{R}^A$; and $C = \bigcup_{a \in A}\{C_a\}$ where, 
for each $a \in A$, $C_a$ is a function: $\Lambda^{\pre^{(A, R, C)}(a)} 
\rightarrow 
\mathcal{L}$. Here and elsewhere, 
$\pre^{(A, R, C)}(a) = \{a_x \in A \mid (a_x, a) \in R\}$ 
for any finite {\ADF} tuple $(A, R, C)$. 
We denote the class of all finite {\ADF} tuples 
by $\FADF$, and refer to its member by $\FFADF$ with or without 
a subscript. 

Moreover, to ease the juxtaposition with \MMA, 
we define the notion of label designation for {\ADF} as well. 
For any $\FFADF \equiv (A, R, C)\ (\in \FADF)$, 
any $a \in A$ and any $\lambda \in \Lambda$ defined 
at least for each member of $\pre^{\FFADF}(a)$, let 
$\lambda\!\!\downarrow_{\pre^{F^{\ADF}}(a)}$ denote
 a member of 
		$\Lambda^{\pre^{F^{\ADF}}(a)}$ with 
$\lambda(a_x) = \lambda\!\!\downarrow_{\pre^{F^{\ADF}}(a)}(a_x)$
		for any $a_x \in \pre^{F^{\ADF}}(a)$. 
Then, for any $\FFADF \equiv (A, R, C) \ (\in \FADF)$, any $a \in A$ 
and any $\lambda \in \Lambda$,  
we say that $\lambda$ designates $l \in \mathcal{L}$ for $a \in A$ iff (1)   
$\lambda$ is defined at least for each member of 
$\pre^{\FFADF}(a)$, and 
	  (2) 
	$C_a(\lambda\!\!\downarrow_{\pre^{F^{\ADF}}(a)})(a) = l$. 
 We say that $a \in A$'s label is proper under $\lambda \in \Lambda$ 
 in $\FFADF$
iff (1) $\lambda$ is defined for $a$, and (2) 
$\lambda$ designates 
$\lambda(a)$ for $a$. 

Since we defined exact labellings of $\MMA$, again to ease 
the juxtaposition, we define it for {\ADF} here. If every $a \in A$'s label is designated 
under $\lambda \in \Lambda^A$, then we say $\lambda$ is an {\it exact labelling} 
of $\FFADF$. Suppose the following 
\begin{tikzcd}[
  column sep=small, row sep=small,inner sep=0pt]  
  \underset{C_p}{a_p}  
	\arrow[r] & 
	\underset{C_q}{a_q} \arrow[l,shift left]  
\end{tikzcd}  
with associated conditions. Assume $C_p([a_q:\inL]^{\lambda}) = \undecL$ and 
$C_p([a_q:l]^{\lambda}) = \inL$ for $l \in \{\undecL, \outL\}$. 
Assume $C_q([a_p:\inL]^{\lambda}) = \undecL$ 
and $C_q([a_p:l]^{\lambda}) = \inL$ for $l \in \{\undecL,\outL\}$. 
Then $[a_p:\inL,a_q:\undecL]^{\lambda}$ and 
$[a_p:\undecL,a_q:\inL]^{\lambda}$ are all the 
exact labellings of $\FFADF$. 
\subsection{Semantics} 
One natural semantics for both {\MMA} and {\ADF} is 
the {\it exact semantics} as the set of all exact labellings.  
However, it is in general not possible to guarantee 
existence of an exact labelling; see a counter-example in \cite{arisaka2020a}.   
While the non-existence is not in itself 
a problem, {\ADF} and {\MMA} both propose 
some approximation, the former with a consensus operator 
and the latter with maximisation of the number of 
arguments whose labels are designated (with the remaining 
labelled $\undecL$), for gaining the existence property. 

Since our objective is to consider properties of {\ADF} 
from within \MMA, it makes sense to touch upon \ADF's consensus operator here, 
and define it also for {\MMA} (which is incidentally 
new; however, being straightforward, we do not have to claim 
any novelty in this formulation for \MMA). Let $\maxi: \mathcal{A} \times  
\Lambda \rightarrow 2^{\Lambda}$, 
which we alternatively 
state $\maxi_{\mathcal{A}}: 
\Lambda \rightarrow 2^{\Lambda}$, 
be such that, 
for any $F \equiv (A, R, X)\ (\in \FADF \cup \FMMA)$, 
any $A \subseteq_{\textsf{fin}} 
\mathcal{A}$ and any 
$\lambda \in \Lambda^A$,  \\\\
\indent $\maxi_A(\lambda) = 
\{\lambda_x \in \Lambda^A \ | \ 
  \lambda \preceq \lambda_x \text{ and } 
\lambda_x \text{ is maximal in } 
(\Lambda^A, \preceq)\}$.\\ 

\noindent Every member $\lambda_x$ of $\maxi_A(\lambda)$ 
is such that $\lambda_x(a) \in \{\inL, \outL\}$ 
for every $a \in A$. Now, 
let $\Theta 
: (\mathcal{F}^{\ADF} \cup \FMMA) \times \Lambda 
\rightarrow \Lambda$, 
which we alternatively 
state $\Theta^{(\mathcal{F}^{\ADF} \cup \FMMA)}: 
\Lambda \rightarrow 
\Lambda$, 
be such that, 
for any $F^{\ADF} \equiv (A, R, C)\  (\in \mathcal{F}^{\ADF})$, 
any $F^{\MMA} \equiv (A, R, f_Q)\ (\in \FMMA)$ 
and any $\lambda \in \Lambda^A$,  all the following hold, 
with $Y$ denoting either of $\FFADF$ and $\FFMMA$. 
\begin{enumerate} 
  \item 
	  $\Theta^{Y}(\lambda)
     \in \Lambda^A$.
   \item For every $a \in A$ and every $l \in \{\inL,\outL\}$, 
	   $\Theta^{Y}(\lambda)(a) = l$ iff, 
		for every 
		$\lambda_x \in \maxi_A(\lambda)$,  $\lambda_x$ 
		designates only $l$ for $a$ in $Y$. 
\end{enumerate}
\noindent In a nutshell \cite{Brewka13}, $\Theta^{Y}(\lambda) 
$ 
gets a consensus of every $\lambda_x \in \maxi_A(\lambda)$ 
on the label of each $a \in A$: if each one of them 
says only $\inL$ for $a$, then $\Theta^{Y}(\lambda)(a) = 
\inL$, 
if each one of them says only $\outL$ for $a$, then 
$\Theta^{Y}(\lambda)(a) = \outL$,
and for the other cases $\Theta^{Y}(\lambda)(a) = \undecL$.  

Then the \ADF-grounded semantics of $Y \in (\FADF \cup \FMMA)$ 
contains just the least fixpoint of $\Theta^{Y}$
(the order is $\preceq$). Readers are referred to \cite{Brewka13} 
for any other \ADF-semantics.

%
%
%
%

\subsection{Abstract interpretation and a Galois connection}  
Abstract interpretation \cite{Cousot77,Cousot79} is a popular 
technique in static analysis, useful for 
reasoning about properties of a large-scale program through abstraction.   
It abstracts a concrete program into an abstract program while 
ensuring that the abstraction be a sound over-approximation 
of the concrete program for some property. The soundness 
is in the sense that if an abstracted program 
satisfies some property, then some property 
is guaranteed to hold in the concrete program. 

Important to abstract interpretation 
is the notion of Galois connection 
(see any standard text, e.g. \cite{Davey02}). 
Briefly, let $S_1$ and $S_2$ each be an ordered set, 
partially ordered in $\leq_1$ and respectively in $\leq_2$. 
Let $f_{1\rightarrow 2}: S_1 \rightarrow S_2$ be a function that maps 
each element of $S_1$ onto an element of $S_2$, 
and let $f_{2 \rightarrow 1}: S_2 \rightarrow S_1$ be a function that maps 
each element of $S_2$ onto an element of $S_1$.  
If $f_{1 \rightarrow 2}(s_1) \leq_2 s_2$ materially implies 
$s_1 \leq_1 f_{2\rightarrow 1}(s_2)$ and vice versa, then 
the pair of $f_{1 \rightarrow 2}$ and $f_{2 \rightarrow 1}$ is said to be a 
Galois connection. The following properties hold good. A Galois connection is: 
contractive, i.e. $(f_{1 \rightarrow 2} \circ f_{2\rightarrow 1})(s_2) \leq_2 s_2$ for every $s_2 \in S_2$; extensive, i.e. $s_1 \leq_1 (f_{2 \rightarrow 1}
\circ f_{1 \rightarrow 2})(s_1)$ for every $s_1 \in S_1$;  and 
monotone for both 
$f_{1 \rightarrow 2}$ and $f_{2 \rightarrow 1}$ (to follow from the contractiveness and the extensiveness). Further, it holds that 
$f_{1 \rightarrow 2} \circ f_{2 \rightarrow 1} \circ f_{1 \rightarrow 2} = 
f_{1 \rightarrow 2}$ and that 
$f_{2 \rightarrow 1} \circ f_{1 \rightarrow 2} \circ f_{2 \rightarrow 1} = 
f_{2 \rightarrow 1}$. 

\section{Galois connection for {\MMA} and {\ADF} and abstract interpretation}   
In this section, we firstly establish a Galois 
connection for {\MMA} and \ADF. \\\\
\noindent \pmb{\FMMA} \textbf{into} \pmb{\FADF}.
Let us begin by defining mappings of $\FFMMA \equiv (A, R, f_Q)\  (\in \FMMA)$ 
onto $\FFADF \equiv (A, R, C)\ (\in \FADF)$. 
For every $a \in A$, $\lambda \in \Lambda$ designates 
at most one member of $\mathcal{L}$ for $a$ in $\FFADF$ 
whereas it may designate more than one member of $\mathcal{L}$ for $a$ in $\FFMMA$.
This difference 
has to be taken into account. 
Example \ref{ex_concretisation} 
shows a concrete mapping example. 
\begin{definition}[Concretisation: from $\FFMMA$ to $F^{\ADF}$]\label{def_concretisation_from} 
	Let $\Gamma$ be a class of 
	all functions $\gamma: \FMMA \rightarrow \FADF$, 
	each of which is 
	such that, for any $\FFMMA \equiv (A, R, f_Q)\ (\in \FMMA)$, 
	$\gamma(\FFMMA)$ is some $(A, R, C) \in \FADF$ 
	with $C$ satisfying the following. For any 
	$a \in A$ and any $\lambda 
			\in \Lambda^{\pre^{\FFMMA}(a)}$, if 
			$L \subseteq \mathcal{L}$ 
	is such that $\lambda$ designates each $l \in L$ but does not 
	designate any $l \in (\mathcal{L} \backslash L)$ for $a$ in 
			$\FFMMA$, then 
	$C_a(\lambda) \in L$.  
	
	We say that $F^{\ADF} \in \mathcal{F}^{\ADF}$ is a concretisation 
	of $\FFMMA \in \FMMA$ iff there is some $\gamma \in \Gamma$ 
	with $F^{\ADF} = \gamma(\FFMMA)$. By $\pmb{\Gamma}[\FFMMA]$ we denote  
	the set of all concretisations of $\FFMMA \in \FMMA$.

	
\end{definition}  
\begin{example}[Concretisation]\label{ex_concretisation} 
	Consider the $\FFMMA \equiv (A, R, f_Q) \ 
	(\in \FMMA)$ in the diagram below. We show some of its concretisations. 
	Since $\gamma \in \Gamma$ does not modify 
	$A$ and $R$, the problem at hand is identification of 
	$C_{a_x}$ to correspond to $f_Q(a_x)$ for each $x \in \{1, \ldots, 5\}$. 
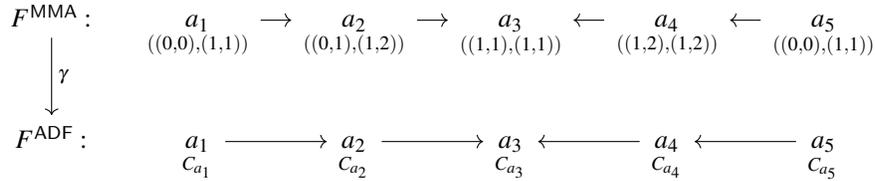
  \begin{center} 
	  \begin{tikzcd}[column sep=small]
		  \FFMMA: \arrow[d,"\gamma"] & 
   \overset{}{\underset{((0, 0), (1,1))}{\ensuremath{a_1}}} 
  \rar & 
\overset{}{\underset{((0,1), (1,2))}{\ensuremath{a_2}}} \rar & 
   \overset{}{\underset{((1,1),(1,1))}{\ensuremath{a_3}}} &  
   \overset{}{\underset{((1,2),(1,2) )}{\ensuremath{a_4}}} \lar &   
   \overset{}{\underset{((0,0), (1, 1))}{\ensuremath{a_5}}} \lar \\
		  F^{\ADF}: & 
		    \overset{}{\underset{C_{a_1}}{\ensuremath{a_1}}} 
  \rar & 
		    \overset{}{\underset{C_{a_2}}{\ensuremath{a_2}}} \rar & 
		    \overset{}{\underset{C_{a_3}}{\ensuremath{a_3}}} &  
		    \overset{}{\underset{C_{a_4}}{\ensuremath{a_4}}} \lar &   
		    \overset{}{\underset{C_{a_5}}{\ensuremath{a_5}}} \lar
\end{tikzcd}   
  \end{center}
	For both $a_1$ and $a_5$, 
	  every $\lambda \in \Lambda$ 
	  designates just $\inL$. 
	  As per \mbox{Definition \ref{def_concretisation_from}},  
	  for any $\lambda \in \Lambda^{\emptyset}$, 
	  $C_{a_1}(\lambda) = C_{a_5}(\lambda) = \inL$ irrespective 
	  of which member of $\Gamma$ is referred to by $\gamma$. 
          
	  For $a_4$, there are 3 distinct labellings $[a_5:\inL]^{\lambda}, 
	   [a_5:\outL]^{\lambda}, [a_5:\undecL]^{\lambda} \in 
	   \Lambda^{\pre^{\FFMMA}(a_4)}$. 
	    We have: $[a_5:\inL]^{\lambda}$ (satisfying not-a and may$_s$-r) 
	    designates 
	    $\outL$ and $\undecL$ for $a_4$; 
	    $[a_5:\outL]^{\lambda}$ (satisfying may$_s$-a and \mbox{not-r}) designates 
	    $\inL$ and $\undecL$ for $a_4$; 
	    and $[a_5:\undecL]^{\lambda}$ (satisfying not-a and not-r) 
	    designates $\undecL$ for $a_4$, in $\FFMMA$.  
	   Thus, $C_{a_4}$ is any one of the following. 
	   \begin{enumerate} 
		   \item $C_{a_4}([a_5:\inL]^{\lambda}) = \outL$, 
			   $C_{a_4}([a_5:\outL]^{\lambda}) = \inL$, 
			   $C_{a_4}([a_5:\undecL]^{\lambda}) = \undecL$.  
		   \item 
			   $C_{a_4}([a_5:\inL]^{\lambda}) = \outL$, 
			   $C_{a_4}([a_5:\outL]^{\lambda}) = \undecL$, 
			   $C_{a_4}([a_5:\undecL]^{\lambda}) = \undecL$. 
		   \item $C_{a_4}([a_5:\inL]^{\lambda}) = \undecL$, 
			   $C_{a_4}([a_5:\outL]^{\lambda}) = \inL$, 
			   $C_{a_4}([a_5:\undecL]^{\lambda}) = \undecL$.   
		   \item 
			   $C_{a_4}([a_5:\inL]^{\lambda}) = \undecL$, 
			   $C_{a_4}([a_5:\outL]^{\lambda}) = \undecL$, 
			   $C_{a_4}([a_5:\undecL]^{\lambda}) = \undecL$.  
	   \end{enumerate}
	   Hence, some $\gamma \in \Gamma$ has the first $C_{a_4}$, 
	   some others have the second, third, or the fourth 
	   $C_{a_4}$. 
           Analogously  for $a_2$ and $a_3$. 
	    \hfill$\clubsuit$       
\end{example} 
\noindent When we either decrease a may- condition ($\in \mathbb{N}$) or 
increase a must- condition of a may-must scale in $\FFMMA \equiv (A, R, f_Q) 
\ (\in \FMMA)$, we 
obtain at least as large a set of concretisations as before the change (\mbox{Theorem \ref{lem_monotonicity_1}}). For the proof, we first read  
the following subsumption relation off Fig. \ref{fig_2}. 
\begin{lemma}[Label designation subsumption]\label{lem_label_designation_subsumption}       
	Let $x, y$ be a member of $\{\text{must}, \text{may}_s, \text{not}\}$. 
	For any $\FFMMA \equiv (A, R, f_Q)\ (\in \FMMA)$, 
	any $a \in A$ and any $\lambda_1, \lambda_2, \lambda_3 \in \Lambda$, 
	if  $a$ satisfies: $x$-a and $y$-r under $\lambda_1$; 
	may$_s$-a and $y$-r under $\lambda_2$; 
	and $x$-a and may$_s$-r under $\lambda_3$, 
	and if $\lambda_1$ designates $l \in \mathcal{L}$ 
	for $a$, then 
	both $\lambda_2$ and $\lambda_3$ designate $l$ 
	for $a$.  
\end{lemma} 

\noindent Also, for convenience, we define the following order.  
\begin{definition}[Abstract order] 
  Let $\unlhd \subseteq \mathcal{Q} \times \mathcal{Q}$ 
	be such that $(Q_1, Q_2) \in \unlhd$, alternatively 
	 $Q_1 \unlhd Q_2$, holds iff, 
	 for any $i \in \{1,2\}$, 
	  $((Q_2)^i)^1 \leq ((Q_1)^i)^1$ and 
	  $((Q_1)^i)^2 \leq ((Q_2)^i)^2$ both hold. 
	  We define $\sqsubseteq\ \subseteq \FMMA \times \FMMA$ 
	  to be such that, 
	  for any $\FFMMA_1 \equiv (A, R, f_Q)$ 
	  and any $\FFMMA_2 \equiv (A, R, f_Q')$, 
	  $(\FFMMA_1, \FFMMA_2) \in\ \sqsubseteq$, alternatively 
	  $\FFMMA_1 \sqsubseteq \FFMMA_2$, holds iff, 
	  for any $a \in A$, 
	  $f_Q(a) \unlhd f_Q'(a)$ holds.  
	  
	  We also extend $\sqsubseteq$ for $2^{\FMMA}$ 
	  in the following manner. 
	  For $\FMMA_x, \FMMA_y \subseteq \FMMA$,  
	  we define: 
	  $\FMMA_x \sqsubseteq \FMMA_y$ 
	  iff, for any $\FFMMA_x \in \FMMA_x$, 
	  there exists some $\FFMMA_y \in \FMMA_y$ 
	  such that $\FFMMA_x \sqsubseteq \FFMMA_y$. 
\end{definition}  

\begin{theorem}[Monotonicity]\label{lem_monotonicity_1} 
	  For any $\FFMMA \equiv (A, R, f_Q)\ (\in \FMMA)$ 
	  and any ${\FFMMA}' \equiv (A, R, f_Q')\ (\in \FMMA)$,  
	  if $\FFMMA \sqsubseteq {\FFMMA}'$ holds, then 
	  $\pmb{\Gamma}[\FFMMA] \subseteq \pmb{\Gamma}[{\FFMMA}']$ holds.  
\end{theorem} 
\textbf{Proof} 
	By Definition \ref{def_concretisation_from}, it 
	 suffices to show that, for any $a \in A$ and 
	any $\lambda \in \Lambda^{\pre^{\FFMMA}(a)}$, if $\lambda$ 
	designates 
	$l \in \mathcal{L}$ for $a$ in $\FFMMA$, then 
	$\lambda$ designates $l$ for $a$ in ${\FFMMA}'$. Now, the differences 
	between $f_Q$ and 
	$f'_Q$ are such that, for any $\lambda \in \Lambda$, firstly, 
	if $a$ satisfies 
	must-a (resp. must-r) under $\lambda$ in $\FFMMA$, $a$ 
	satisfies either must-a 
	or may$_s$-a (resp. must-r or may$_s$-r) under $\lambda$ in 
	${\FFMMA}'$, and, secondly, if $a$ satisfies not-a 
	(resp. not-r) under $\lambda$ in $\FFMMA$, $a$ 
	satisfies either not-a or may$_s$-a 
	(resp. not-r or may$_s$-r) under $\lambda$ in ${\FFMMA}'$. 
	Apply 
	Lemma \ref{lem_label_designation_subsumption}. \hfill$\Box$ \\

\noindent \pmb{\FADF} \textbf{into} \pmb{\FMMA}. Into the other direction of mapping $F^{\ADF} \equiv (A, R, C)\ 
(\in \mathcal{F}^{\ADF})$ 
onto $\FFMMA \equiv (A, R, f_Q)\ (\in \FMMA)$, recall that 
$\FFMMA$ only requires $a\ (\in A)$'s may-must scales, 
$|\pre^{\FFMMA}_{\lambda, \inL}(a)|$ 
and $|\pre^{\FFMMA}_{\lambda, \outL}(a)|$ (for $\lambda \in \Lambda$) for label designation. 
For any $\lambda_1, \lambda_2 \in \Lambda$, 
as long as $a$ satisfies $x$-a and $y$-r ($x, y \in \{\text{must}, 
\text{may}_s, \text{not}\}$) under both $\lambda_1, \lambda_2$, 
it holds that 
$\lambda_1$ and $\lambda_2$ designate the same label(s) for $a$. 

On the other hand (see also \mbox{Example 
\ref{ex_concretisation}}), 
$F^{\ADF}$'s $C_a$ determines label designation independently for each 
$\lambda \in \Lambda^{\pre^{F^{\ADF}}(a)}$. 
For distinct $\lambda_1, \lambda_2 \in \Lambda^{\pre^{F^{\ADF}}(a)}$ 
with $|\pre^{\FFADF}_{\lambda_1, \inL}(a)| = |\pre^{\FFADF}_{\lambda_2, \inL}(a)|$ 
and $|\pre^{\FFADF}_{\lambda_1, \outL}(a)| = |\pre^{\FFADF}_{\lambda_2, 
\outL}(a)|$,  it can happen that $C_a(\lambda_1) \not= C_a(\lambda_2)$.

As such, we need to abstract the specificity of $F^{\ADF}$'s $C$ 
 for the mapping. 
Formally, we consider the following class 
of functions.  
See Example \ref{ex_abstraction} to follow 
for a concrete example. 
\begin{definition}[Abstraction: from $\FFADF$ to $\FFMMA$] \label{def_abstraction_from} 
	  Let $\Delta$ be a class of all functions $\alpha: 
	 \mathcal{F}^{\ADF} \rightarrow \FFMMA$, 
	 each of which is such that, for any 
	  $F^{\ADF} \equiv (A, R, C)\ (\in \mathcal{F}^{\ADF})$, 
	 $\alpha(F^{\ADF})$ is some $(A, R, f_Q) \in \FMMA$
	 where, for every $a \in A$ and every $\lambda \in \Lambda^{\pre^{\FFADF}(a)}$, 
	 $f_Q(a) \equiv ((n_1^a, n_2^a), (m_1^a, m_2^a))$ 
	  satisfies all the following 
	  conditions. 
                       \begin{enumerate}  
			       \item $0 \leq n^a_1 \leq n^a_2 \leq |\pre^{\FFADF}(a)|+1$. 
				       Also 
				       $0 \leq m^a_1 \leq m^a_2 \leq 
				        |\pre^{\FFADF}(a)| + 1$.\\                             
			       \item If $|\pre^{\FFADF}_{\lambda, \outL}{(a)}| 
				       < n_1^{a}$ and 
				       $|\pre^{\FFADF}_{\lambda, \inL}{(a)}| 
				       < m_1^{a}$, then  
				       $C_{a}(\lambda) = \undecL$.\\
				       (This corresponds to \textbf{not-a, not-r} 
				       satisfaction.)\\
			       \item
				       If $n_1^{a} \leq 
				       |\pre^{\FFADF}_{\lambda, \outL}{(a)}| 
				       < n_2^{a}$ and 
				       $|\pre^{\FFADF}_{\lambda, \inL}{(a)}| 
				       < m_1^{a}$, then  
				       $C_{a}(\lambda) \in \{\inL, \undecL\}$. \\
				       (\textbf{may$_s$-a, not-r})  \\
			       \item
				       If $n_2^{a} \leq 
				       |\pre^{\FFADF}_{\lambda, \outL}{(a)}|$
				       and 
				       $|\pre^{\FFADF}_{\lambda, \inL}{(a)}| 
				       < m_1^{a}$, then  
				       $C_{a}(\lambda) = \inL$. \\
				        (\textbf{must-a, not-r}) \\ 
			       \item If $|\pre^{\FFADF}_{\lambda, \outL}{(a)}| 
				       < n_1^{a}$ and 
				       $m_1^{a} \leq 
				       |\pre^{\FFADF}_{\lambda, \inL}{(a)}| 
				       < m_2^{a}$, then 
				       $C_{a}(\lambda) \in 
				       \{\outL, \undecL\}$.\\
				        (\textbf{not-a, may$_s$-r})\\
			       \item
				       If $n_1^{a} \leq 
				       |\pre^{\FFADF}_{\lambda, \outL}{(a)}| 
				       < n_2^{a}$ and  
				       $m_1^{a} \leq 
				       |\pre^{\FFADF}_{\lambda, \inL}{(a)}| 
				       < m_2^{a}$,
				         then  
				       $C_{a}(\lambda) 
				       \in \{\inL,\outL,\undecL\}$.\\
				        (\textbf{may$_s$-a, may$_s$-r})\\ 
			       \item
				       If $n_2^{a} \leq 
				       |\pre^{\FFADF}_{\lambda, \outL}{(a)}|$
				       and  
				       $m_1^{a} \leq 
				       |\pre^{\FFADF}_{\lambda, \inL}{(a)}| 
				       < m_2^{a}$,
				        then  
				       $C_{a}(\lambda) \in \{\inL, \undecL\}$. \\
				       (\textbf{must-a, may$_s$-r}) \\ 
			       \item
				      If $|\pre^{\FFADF}_{\lambda, \outL}{(a)}| 
				       < n_1^{a}$ and 
				       $m_2^{a} \leq 
				       |\pre^{\FFADF}_{\lambda, \inL}{(a)}| 
				       $, then 
				       $C_{a}(\lambda) = \outL$. \\
				       (\textbf{not-a, must-r})\\
			       \item
				       If $n_1^{a} \leq 
				       |\pre^{\FFADF}_{\lambda, \outL}{(a)}| 
				       < n_2^{a}$ and  
				       $m_2^{a} \leq 
				       |\pre^{\FFADF}_{\lambda, \inL}{(a)}| 
				       $,
				         then  
				       $C_{a}(\lambda) \in \{\outL,\undecL\}$. \\
				        (\textbf{may$_s$-a, must-r})\\ 
			       \item
				       If $n_2^{a} \leq 
				       |\pre^{\FFADF}_{\lambda, \outL}{(a)}|$
				       and  
				       $m_2^{a} \leq 
				       |\pre^{\FFADF}_{\lambda, \inL}{(a)}| 
				       $, 
				        then  
				       $C_{a}(\lambda) = \undecL$.\\
				        (\textbf{must-a, must-r}) 
		       \end{enumerate} 
	For any $F^{\ADF} \in \mathcal{F}^{\ADF}$,  
	we say that $\FFMMA \in \FMMA$ is an abstraction of $F^{\ADF}
	$ iff there is some $\alpha \in \Delta$ 
	with $\FFMMA = \alpha(F^{\ADF})$. We 
        denote the set of all abstractions of $F^{\ADF}$ 
	by $\pmb{\Delta}[F^{\ADF}]$. 
	\end{definition}

\begin{example}[Abstraction] \label{ex_abstraction}  
	Suppose $F^{\ADF}:$ \begin{tikzcd}[column sep=small]
		{\underset{C_{a_1}}{\ensuremath{a_1}}} 
  \rar & 
		{\underset{C_{a_2}}{\ensuremath{a_2}}}  & 
		\overset{}{\underset{C_{a_3}}{\ensuremath{a_3}}} \lar . 
   \end{tikzcd}   
	Then $\Lambda^{\pre^{F^{\ADF}}(a_2)} = 
	\bigcup_{l_1, l_3 \in \mathcal{L}}\{[a_1:l_1,a_3:l_3]^{\lambda}\}$. 
	Assume: 
	\begin{enumerate} 
		\item $C_{a_2}([a_1:\undecL,a_3:\undecL]^{\lambda}) = \undecL$.  
		  \quad
		   {\normalfont 2.} \ 
			$C_{a_2}([a_1:l_1,a_3:l_3]^{\lambda}) = \outL$ if $\inL \in 
		   \{l_1\} \cup \{l_3\}$.  
			\setcounter{enumi}{2}
		\item $C_{a_2}([a_1:\undecL,a_3:\outL]^{\lambda}) = \inL$. \qquad\quad\ \
		   {\normalfont 4.} \ 
			$C_{a_2}([a_1:\outL,a_3:\undecL]^{\lambda}) = \undecL$. 
			\setcounter{enumi}{4}
		\item	$C_{a_2}([a_1:\outL,a_3:\outL]^{\lambda}) = \undecL$. 
	\end{enumerate} 	
	Let us consider which $f_Q(a_2) \equiv ((n_1^{a_2}, n_2^{a_2}), 
	(m_1^{a_2}, m_2^{a_2}))$ 
	can be 
	in an abstraction of $F^{\ADF}$.
	Firstly, by 1. and 2., for any  
	$\lambda \in \Lambda$ defined for $a_1$ and $a_3$, 
	we see that 
	$\lambda$ does not designate $\outL$ for $a_2$ so long as 
	$|\pre^{\FFADF}_{\lambda, \inL}(a_2)| = 0$; however, 
	as soon as $|\pre^{\FFADF}_{\lambda, \inL}(a_2)| > 0$, 
	we have that $\lambda$ designates only $\outL$. 
	As the result, we can set $(m_1^{a_2}, m_2^{a_2})$ 
	to $(1, 1)$. On the other hand, for $(n_1^{a_2}, n_2^{a_2})$, 
	we have 3. and 4., and it cannot be that $n_2^{a_2} \leq 1$, 
	but also we have 5., and $n_2^{a_2} \not=2$. Thus, 
	we must have $n_2^{a_2} = 3$. However, because of 3., 
	$n_1^{a_2}$ cannot be greater than or equal to 2. 
	Hence we can set $n_1^{a_2}$ to be 1, resulting 
	in $(n_1^{a_2}, n_2^{a_2}) = (1, 3)$. 
	It is trivial to see that any $\FFMMA \in \pmb{\Delta}[F^{\ADF}]$ 
	with this $f_Q(a_2)$ least abstracts $C_{a_2}$ of $\FFADF$ 
	among all possible $f_Q(a_2)$ in members of $\pmb{\Delta}[F^{\ADF}]$. 
	Other $f_Q(a_2)$ also appear in other members of 
	$\pmb{\Delta}[F^{\ADF}]$. 
	Specifically, due to Lemma \ref{lem_label_designation_subsumption} 
	and Definition \ref{def_abstraction_from}, 
	$\FFMMA \in \pmb{\Delta}[F^{\ADF}]$ 
        can come with 
	any $f_Q(a_2) = (({n_1^{a_2}}', 3), ({m_1^{a_2}}', {m_2^{a_2}}'))$ 
	with $0 \leq {n_1^{a_2}}' \leq 1$,  $0 \leq {m_1^{a_2}}' \leq 1$, 
	$1 \leq {m_2^{a_2}}' \leq 3$. 
	Analogously for $f_Q(a_1)$ and $f_Q(a_3)$.  
	\hfill$\clubsuit$ 
\end{example} 
Until now, we have left the cardinality of $\pmb{\Gamma}[\FFMMA]$ and 
$\pmb{\Delta}[F^{\ADF}]$  all up to intuition. The following 
theorem establishes the bounds, with an implication of 
existence of the two sets (Corollary \ref{cor_existence}).  
\begin{theorem}[Cardinality of the maps]  
	For any $\FFMMA \equiv (A, R, f_Q)\ (\in \FMMA)$,  
	we have 
	$1 \leq \pmb{\Gamma}[\FFMMA] \leq |\mathcal{L}|^{|A|(|\mathcal{L}|^{|A|})}$, 
	and for 
	any $F^{\ADF} \equiv (A, R, C)\ (\in {\mathcal{F}^{\ADF}})$,  
	we have $1 \leq \pmb{\Delta}[F^{\ADF}]
	\leq (\frac{(|A|+2)(|A|+3)}{2})^{2|A|}$. 
	It holds that \linebreak
	$(\frac{(|A|+2)(|A|+3)}{2})^{2|A|} \ll 
	|\mathcal{L}|^{|A|(|\mathcal{L}|^{|A|})}$ for any $2 \leq |A|$ 
	and $|\mathcal{L}|=3$. 
\end{theorem}  
\textbf{Proof.} 
	For the former, the lower bound is due to the fact that any 
	$\lambda \in \Lambda^{\pre^{\FFMMA}(a)}$ for $a \in A$ 
	designates at least one $l \in \mathcal{L}$ for $a$ in $\FFMMA$.  
	For the upper bound, $|\pre^{\FFMMA}(a)| \leq |A|$, and thus 
	$|\Lambda^{\pre^{\FFMMA}(a)}| \leq |\mathcal{L}|^{|A|}$. 
	Any $\lambda \in \Lambda^{\pre^{\FFMMA}(a)}$  
	may designate as many labels as there are in $\mathcal{L}$  
	for $a$, hence there may be up to $|\mathcal{L}|^{(|\mathcal{L}|^{|A|})}$
	alternatives for the third component of 
	$F^{\ADF} \in \pmb{\Gamma}(\FFMMA)$. There are $|A|$ arguments. 
	Put together, we obtain the result.  
	For the latter,  
	the lower bound is due to the fact that 
	 $f_Q(a) = ((0, |\pre^{\FFMMA}(a)|+1), (0, |\pre^{\FFMMA}(a)|+1))$  
	 designates each of $\inL, \outL$ and $\undecL$. 
        For the upper bound, 
	for each $a \in A$,  
	$C_a$ 
	 may map to 
	member(s) of 
	$X \equiv \{((n_1,n_2), (m_1,m_2)) \mid 0 \leq n_1, n_2, m_1, m_2\leq |\pre^{F}(a)| + 1\}$. 
Since we have $|\pre^{\FFMMA}(a)| + 1 \leq |A|+1$, it holds that 
	$|X| \leq (\frac{(|A|+2)(|A|+3)}{2})^2$. There are $|A|$ arguments.  
		 \hfill$\Box$ \\
\begin{corollary}[Existence]\label{cor_existence} 
	For any $\FFMMA \in \FMMA$, 
	there exists some concretisation 
	of $\FFMMA$, and 
	for any $\FFADF \in \FADF$, 
	there exists some abstraction of 
	$\FFADF$. 
\end{corollary} 
While $\pmb{\Delta}[\FFADF]$ 
is still rather large, note that it covers every possible abstraction
of $\FFADF$. In practice, 
we adopt only a single set of criteria for abstraction  
e.g. minimal abstraction (see $f_{\alpha}$ in Theorem \ref{thm_galois_connection}) 
in $\sqsubseteq$, 
which leaves only a handful of the set, or just one 
in case the abstraction minimum in $\sqsubseteq$ exists. 
 
Any abstraction of $F^{\ADF} \in {\mathcal{F}^{\ADF}}$ 
correctly overapproximates $C_a$'s label 
designation for every $a \in A$, following trivially from   
the definition of label designation (Section 2) 
 and 
Definition \ref{def_abstraction_from}. 
\begin{proposition}[Abstraction soundness]\label{prop_abstraction_soundness} 
	For any $F^{\ADF} \equiv (A, R, C)\ (\in {\mathcal{F}^{\ADF}})$ and 
	any $\FFMMA \equiv (A, R, f_Q)\ (\in \FMMA)$, 
	if $\FFMMA$ is an abstraction of $\FFADF$, 
	then for any $a \in A$ and any $\lambda \in \Lambda^{\pre^{\FFMMA}(a)}$
	all the following hold. 
	\begin{enumerate} 
   		\item $\lambda$ designates $C_a(\lambda)$ for $a$ in $\FFMMA$.  
		\item if $\lambda$ designates at most one $l \in \mathcal{L}$  
			  for $a$ in $\FFMMA$, then 
			$\lambda$ designates $l$ for $a$ in $\FFADF$. 
	\end{enumerate}
\end{proposition} 
\subsection{Galois connection}  
When there are two systems related in concrete-abstract relation, 
it is of interest to establish Galois connection (Cf. Section 2). 
Galois connection is used for 
abstraction interpretation 
in static analysis 
for verification of properties of 
large-scale programs as the verification in concrete space
is often undecidable or very costly. 
In our view, it is no different with formal argumentation; 
reasoning  
about a large-scale argumentation
will benefit from utilising the technique. 
Here, we identify a Galois connection 
between $\FMMA$ and $\mathcal{F}^{\ADF}$ 
based on $\Gamma$ and $\Delta$ that we introduced. \pagebreak
\begin{theorem}[Galois connection] \label{thm_galois_connection}  
	Let $2^{\FMMA}_{(A, R)}$  
	be a subclass of $2^{\FMMA}$ 
	which contains every $\FFMMA \equiv (A, R, f_Q)\ (\in \FMMA)$ 
	for some $f_Q$ but nothing else.
	Let $2^{\mathcal{F}^{\ADF}}_{(A, R)}$  
	be a subclass of $2^{\mathcal{F}^{\ADF}}$ 
	which contains every $F^{\ADF} \equiv (A, R, C)\ (\in \mathcal{F}^{\ADF})$ 
	for some $C$ but nothing else.

	Let $f_{\gamma}: 2^{\FMMA} \rightarrow 2^{\FADF}$ 
	and  $f_{\alpha}: 2^{\FADF} \rightarrow 2^{\FMMA}$ 
        be the following.
	\begin{itemize} 
		\item for any $\FMMA_x \in 2^{\FMMA}_{(A, R)}$, 
	$f_{\gamma}(\FMMA_x) = 
	\bigcup_{\FFMMA_x \in \FMMA_x}\pmb{\Gamma}[\FFMMA_x]$.  
\item  for any  $\mathcal{F}^{\ADF}_x \in 2^{\mathcal{F}^{\ADF}}_{(A, R)}$,
	 $f_{\alpha}(\mathcal{F}^{\ADF}_x) =  
			\bigcup_{F_x^{\ADF} \in \mathcal{F}_x^{\ADF}}\{\FFMMA 
			\in \pmb{\Delta}[F_x^{\ADF}] 
			\mid \forall {\FFMMA}' \in \pmb{\Delta}[F_x^{\ADF}].\linebreak
			{\FFMMA}' \not\sqsubset \FFMMA\}$.  
	\end{itemize} 
	 Then 
	 $(f_{\alpha}, f_{\gamma})$ is a Galois connection for 
	 $(2^{\FMMA}_{(A, R)}, \sqsubseteq)$ and 
	 $(2^{\mathcal{F}^{\ADF}}_{(A,R)}, \subseteq)$. 
\end{theorem}
\textbf{Proof.} 
	Suppose $\FMMA_x \in 2^{\FMMA}_{(A, R)}$ and 
	$\mathcal{F}^{\ADF}_x
	 \in 2^{\mathcal{F}^{\ADF}}_{(A, R)}$. Suppose  
	$f_{\alpha}(\mathcal{F}^{\ADF}_x) \sqsubseteq \FMMA_x$, then 
	we have to show that $\mathcal{F}^{\ADF}_x \subseteq  
	f_{\gamma}(\FMMA_x)$. By Theorem \ref{lem_monotonicity_1}, 
	we have $\bigcup_{\FFMMA_y \in f_{\alpha}(\mathcal{F}^{\ADF}_x)}\pmb{\Gamma}[\FFMMA_y] 
	\subseteq 
	\bigcup_{\FFMMA_x \in \FMMA_x}\pmb{\Gamma}[\FMMA_x]$. 
	By the definition of $f_{\gamma}$, 
	we have $\bigcup_{\FFMMA_x \in \FMMA_x}\pmb{\Gamma}[\FFMMA_x] \subseteq  
	f_{\gamma}(\FMMA_x)$. By \mbox{Proposition \ref{prop_abstraction_soundness}} 
	and \mbox{Definition \ref{def_concretisation_from}}, 
	we have $\mathcal{F}_x^{\ADF} \subseteq 
	\bigcup_{\FFMMA_y \in f_{\alpha}(\mathcal{F}^{\ADF}_x)}\pmb{\Gamma}[\FFMMA_y]$. 
	Hence\linebreak $\mathcal{F}_x^{\ADF} \subseteq f_{\gamma}(\FMMA_x)$, as required.

	Suppose on the other hand that 
	$\mathcal{F}^{\ADF}_x \subseteq f_{\gamma}(\FMMA_x)$, then 
	we have to show that 
	$f_{\alpha}(\mathcal{F}^{\ADF}_x) \sqsubseteq 
	\FMMA_x$. First, trivially,  
	   we have $f_{\alpha}(\mathcal{F}_x^{\ADF}) \sqsubseteq 
	  f_{\alpha}(f_{\gamma}(\FMMA_x))$, since 
	  every $F^{\ADF} \in \mathcal{F}_x^{\ADF}$ 
	  is contained in $f_{\gamma}(\FMMA_x)$ by the present assumption. 
	  By Lemma \ref{lem_label_designation_subsumption}, 
	   we have $f_{\alpha}(f_{\gamma}(\FMMA_x)) \sqsubseteq \FMMA_x$, thus 
	  we have $f_{\alpha}(\mathcal{F}_x^{\ADF}) \sqsubseteq \FMMA_x$, 
	  as required. \hfill$\Box$ \\

\subsection{Abstract interpretation on semantics}   
We close this section by presenting that the Galois connection allows 
us to infer semantic properties of $\FFADF \in \FADF$
from within $\FMMA$, for both the exact semantics 
and the \ADF-grounded semantics (Cf. Section 2). We assume $\Label: \{\exact, \ADFg\} \times 
(\FADF \cup \FMMA) \rightarrow 2^{\Lambda}$ to be such that, 
for any $Y \in (\FADF \cup \FMMA)$, 
$\Label(\exact, Y)$ is the exact semantics of $Y$ 
and $\Label(\ADFg, Y)$ is the \ADF-grounded semantics of $Y$. 

\begin{theorem}[Abstract interpretation (exact semantics)]\label{thm_1} 
    For any 
	$\FFMMA \equiv (A, R, f_Q)\ (\in \FMMA)$, 
	let $\Label_{\textsf{one}}(\exact, \FFMMA)$ denote 
	the largest subset of \linebreak
	$\Label(\exact, \FFMMA)$ satisfying: if 
	$\lambda \in \Label_{\textsf{one}}(\exact, \FFMMA)$, 
	then, for every $a \in A$, $\lambda$ designates at most one 
	$l \in \mathcal{L}$ 
	for $a$ in $\FFMMA$. 
	
	Then, 
	for any $F^{\ADF} \equiv (A, R, C) \ (\in \mathcal{F}^{\ADF})$ and any 
	$\FFMMA \in f_{\alpha}(\{\FADF\})$,
	it holds that\linebreak $\Label_{\textsf{one}}(\exact, \FFMMA) \subseteq 
	\Label(\exact, \FFADF)$. 
\end{theorem} 
\textbf{Proof.} Follows from 2. of Proposition \ref{prop_abstraction_soundness}.
\hfill$\Box$\\
\begin{corollary} 
      For any $F^{\ADF} \equiv (A, R, C) \ (\in \mathcal{F}^{\ADF})$ 
	and any $\FFMMA \in f_{\alpha}(\{\FADF\})$, 
	if $\Label_{\textsf{one}}(\exact, \FFMMA) \not= \emptyset$, 
	then $\Label(\exact, \FFADF) \not= \emptyset$. 
\end{corollary}

\begin{theorem}[Abstract interpretation (\ADF-grounded semantics)]\label{thm_2}
    For any $F^{\ADF} \equiv (A, R, C) \ (\in \mathcal{F}^{\ADF})$ and any 
	$\FFMMA \in f_{\alpha}(\{\FADF\})$, 
	it holds for any\linebreak $\lambda \in \Label(\ADFg, \FFMMA)$ that 
	there exists some $\lambda' \in \Label(\ADFg, \FFADF)$ 
	with $\lambda \preceq \lambda'$. 
\end{theorem} 
\textbf{Proof.} Follows from 2. of Proposition \ref{prop_abstraction_soundness}.
\hfill$\Box$ \\
 
In the literature of formal argumentation, 
two types of acceptance (and rejection) of an argument are popularly 
referred to. In the context of labelling-based argumentations,  
for any $(A, R, C) \in \FADF$ or $(A, R, f_Q) \in \FMMA$, 
$a \in A$ is called: credulously accepted (resp. rejected)  
with respect to a semantics 
iff there exists at least one member $\lambda \in \Lambda^A$ of the semantics  
with $\lambda(a) = \inL$ (resp. $\lambda(a) = \outL$); 
and skeptically accepted (resp. rejected)  
with respect to a semantics 
iff $a$ is credulously accepted 
$\andC$ 
$\lambda(a) = \inL$ (resp. $\lambda(a) = \outL$)  
for every member $\lambda \in \Lambda^A$ of the semantics. 

\begin{theorem}[Acceptance and rejection] 
       With respect to both exact and $\ADF$-grounded semantics, 
	for any $\FFADF \equiv (A, R, C)\ (\in \FADF)$, 
	any $\FFMMA \in f_{\alpha}(\{\FADF\})$ and 
	any $a \in A$, all the following hold good. 
        \begin{enumerate} 
		\item if $a$ is credulously accepted (resp. rejected) in $\FFMMA$ 
			with respect to the exact or the $\ADF$-grounded semantics, 
			then $a$ is credulously accepted (resp. rejected) 
			in $\FFADF$ with respect to the same semantics. 
		\item if $a$ is skeptically accepted (resp. rejected) in 
			$\FFMMA$ with respect to the exact semantics 
			and if  $|\Label_{\textsf{one}}(\exact, \FFMMA)|
			= |\Label(\exact, \FFADF)|$, then 
			$a$ is skeptically accepted (resp. rejected) 
			in $\FFADF$ with respect to the exact semantics.  
		\item if $a$ is skeptically accepted (resp. rejected) 
			in $\FFMMA$ with respect to the $\ADF$-grounded semantics,
			then $a$ is skeptically accepted (resp. rejected) 
			in $\FFADF$ with respect to the $\ADF$-grounded semantics.
	\end{enumerate}
\end{theorem} 
\textbf{Proof.} Follows from Theorem \ref{thm_1} and Theorem \ref{thm_2}. 
\hfill$\Box$ 

\section{Conclusion}
We identified a Galois connection 
for abstract dialectical frameworks and 
may-must argumentation, demonstrating  
abstract interpretation at play in formal argumentation. 
The technique of 
abstract interpretation or its significance 
has almost not been known to the argumentation 
community. As far as we are aware, there is a preprint 
\cite{ArisakaDauphin18} that contemplates its application to 
abstraction of 
loops in an argumentation graph to sharpen  
acceptability statuses of $\undecL$-labelled arguments. 
However, while, to an extent, it carries an underlying 
motivation of abstract interpretation 
to know better what could not be otherwise known, 
it is not entirely clear whether the abstraction of loops 
intends a sound over-approximation. 
In comparison, we identified a Galois connection 
between abstract dialectical frameworks and may-must argumentations 
with results stipulating what the semantic predictions in {\MMA} space 
mean in {\ADF} space.

For future work, 
%
it is possible to make the situation 
more complex with concurrency in multi-agent argumentation
\cite{ArisakaSatoh18,arisaka2020b}. Complication is unbounded. 

In static analysis, 
abstract interpretation generally involves 
consideration for widening and narrowing operations.  
Studies on more algorithmic approaches with 
them should be also interesting and practically worthwhile.

\bibliographystyle{acmtrans}
\bibliography{referencesShorter}

\label{lastpage}
\end{document}